\documentclass[aps,prl,twocolumn,superscriptaddress,floatfix,amsmath,showpacs]{revtex4-1}
\usepackage[breaklinks=true,colorlinks=true,linkcolor=blue,urlcolor=blue,citecolor=blue]{hyperref}
\usepackage{graphicx}
\usepackage{dcolumn}
\usepackage{bm}
\usepackage{amssymb}
\usepackage[T1]{fontenc}
\usepackage{textcomp}
\usepackage[]{units}
\usepackage{pgfplots}
\pgfplotsset{compat=newest} 
\pgfplotsset{plot coordinates/math parser=false}
\usepackage{pgfplots}
\usetikzlibrary{pgfplots.groupplots}
\pgfplotsset{compat=1.3}
\usepackage{bm,graphicx,graphics,amsmath,amssymb,bm,epsfig,color}
\usepackage{euscript,tabularx}
\usepackage{longtable}
\usepackage{MnSymbol,wasysym,stmaryrd,ifsym,rotating,marvosym}
\usepackage{xparse}
\usepackage{commath}
\usepackage{changes}

\def\vec#1{{\bm{#1}}}

\def\t#1{{\mathrm{#1}}}

\begin{document}

\title{Simultaneous multitone microwave emission by DC-driven spintronic nano-element}


\author{A. Hamadeh} \affiliation{Fachbereich Physik and Landesforschungszentrum OPTIMAS, Technische Universit\"at Kaiserslautern, 67663 Kaiserslautern, Germany}

\author{D. Slobodianiuk} 
\affiliation{Taras Shevchenko National University of Kyiv, Kyiv 01601, Ukraine}
\affiliation{Institute of Magnetism, Kyiv 03142, Ukraine}

\author{R. Moukhader}\affiliation{Dept. Mathematical and Computer Sciences, Physical Sciences and Earth Sciences, University of Messina, 98166 Messina, Italy}

\author{G. Melkov} \affiliation{Taras Shevchenko National University of Kyiv, Kyiv 01601, Ukraine}

\author{V. Borynskyi} \affiliation{Institute of Magnetism, Kyiv 03142, Ukraine}

\author{M. Mohseni}\affiliation{Fachbereich Physik and Landesforschungszentrum OPTIMAS, Technische Universit\"at Kaiserslautern, 67663 Kaiserslautern, Germany}  

\author{G. Finocchio}\affiliation{Dept. Mathematical and Computer Sciences, Physical Sciences and Earth Sciences, University of Messina, 98166 Messina, Italy}
\author{V. Lomakin}\affiliation{Center for Magnetic Recording Research, University of California San Diego, La Jolla, California 92093-0401, USA}  

\author{R. Verba} \affiliation{Institute of Magnetism, Kyiv 03142, Ukraine}
\author{G. de Loubens} \affiliation{SPEC, CEA, CNRS, Universit\'e Paris-Saclay, 91191 Gif-sur-Yvette, France}
\author{P. Pirro}\affiliation{Fachbereich Physik and Landesforschungszentrum OPTIMAS, Technische Universit\"at Kaiserslautern, 67663 Kaiserslautern, Germany}  
\author{O. Klein} \affiliation{Univ. Grenoble Alpes, CEA, CNRS, Grenoble INP, INAC-Spintec, 38054 Grenoble, France}

\date{\today}

\begin{abstract}
Current-induced self-sustained magnetization oscillations in spin-torque nano-oscillators (STNOs) are promising candidates for ultra-agile microwave sources or detectors. While usually STNOs behave as a monochrome source, we report here clear bimodal simultaneous emission of incommensurate microwave oscillations, where the two tones correspond to two parametrically coupled eigenmodes with tunable splitting. The emission range is crucially sensitive to the change in hybridization of the eigenmodes of free and fixed layers, for instance, through  a slight tilt of the applied magnetic field from the normal of the nano-pillar. 
Our experimental findings are supported both analytically and by micromagnetic simulations, which ascribe the process to four-magnon scattering between a pair of radially symmetric magnon modes and a pair of magnon modes with opposite azimuthal index. Our findings open up new possibilities for cognitive telecommunications and neuromorphic systems that use frequency multiplexing to improve communication performance.
\end{abstract}


\maketitle

A great number of research projects has been devoted to the study of spin transfer torque (STT) after its theoretical prediction \cite{slonczewski1996current, berger1996emission}. This new paradigm is meant to ignite a conceptual metamorphosis of spintronics, a research field which capitalizes on the spin degree of freedom of the electron \cite{wolf2001spintronics}.
The STT effect can enable a variety of spintronics applications, such as spin-torque magnetic random access memories (MRAM) \cite{thomas2014perpendicular} and spin-torque nano-oscillators (STNOs) \cite{hirohata2020review}. The use of STNOs to generate microwave signals in nanoscale devices has generated tremendous and continuous research interest in recent years \cite{kiselev03,mistral06, houssameddine2008spin, mistral08,zeng2013spin,omasello2022antiferromagnetic}. Their key features are frequency tunability \cite{bonetti2009spin}, nanoscale size \cite{zeng2013spin}, broad working temperature \cite{prokopenko2012influence}, and easy integration with the standard silicon technology \cite{kiselev03,rippard2004direct}. As strongly nonlinear devices, STNOs can exhibit different dynamic regimes, which are promising candidates for various applications including microwave signal generation and detection \cite{grimaldi2014spintronic, louis2017low,litvinenko20}, signal modulation \cite{skowronski2019microwave}, spin wave generation \cite{madami2011direct}, neuromorphic computing \cite{torrejon2017neuromorphic, grollier2016spintronic, riou2019temporal}, etc.

Generally, only a $\emph{single}$ mode is expected to oscillate at one time in an STNO as predicted by the universal oscillator model \cite{slavin2009nonlinear}. Multimodal co-generation of weak commensurate tones can be produced by harmonic distortion. Here, the tones are intrinsically linked by rational numbers. Alternatively, mode hopping between nearly degenerate eigen-solutions have been reported  \cite{Eklund13,heinonen13,Iacocca14} induced by thermal fluctuations \cite{Slobodianiuk_CMP2014}, spatial inhomogeneity of the internal field in asymmetric ferromagnetic bilayers \cite{chen2022spatial}, or by formation of higher-order modes of excited magnetic solitons \cite{Yang_SR2015}. Also, high level of thermal fluctuations can results in a seemingly multimode generation that, in fact, is just an amplification of incoherent thermal population of higher modes.  

In this letter, we find and elucidate another possibility to create stable \emph{simultaneous} excitation of multi spin-wave modes in an STO with a continuously adjustable splitting. The leading order mechanism supporting multimode generation is found to be four-magnon scattering or, in other words, second-order parametric instability. Although this process is well-known in magnetic systems \cite{Lvov_Book1994,suhl1957theory,Pirro2021CoherentMagnonics,Pirro2014fourmagnon}, its observation in an STNO is often prohibited as it is impossible to satisfy both energy and angular momentum conservation rules. We found parametric instability becomes possible because of strong hybridization between the modes of the two layers of the STNO, which also makes the parametric process highly sensitive to external conditions and thus controllable.

 \begin{figure}
  \includegraphics[width=\columnwidth]{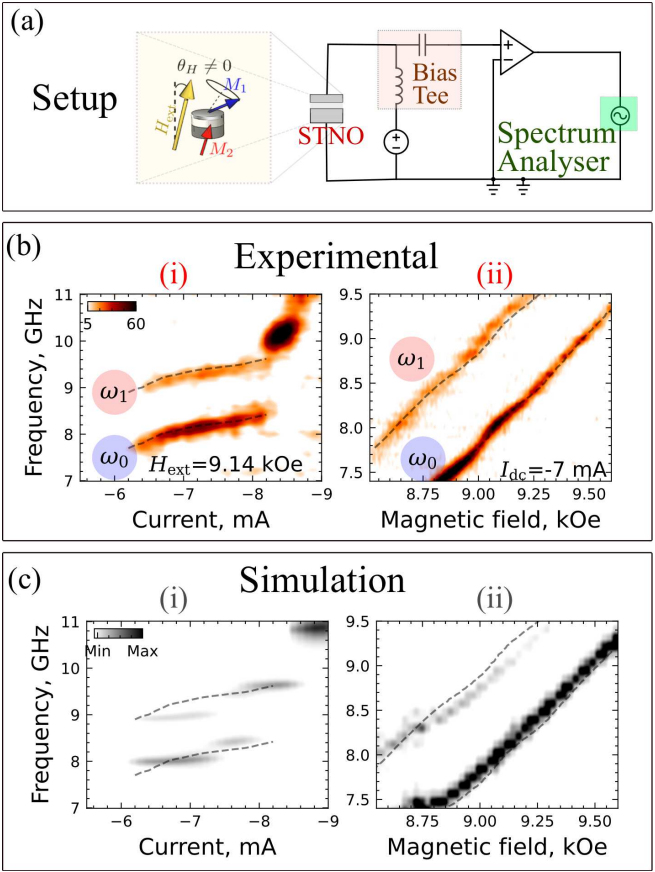}
  \caption{(a) Experimental setup for electrical measurement. A dc current is passed through the STO, while the ac component is extracted via a bias-tee, amplified and measured with a spectrum analyzer. (b,d) Dependence of the microwave generation characteristics of the STNO on the dc current at constant bias field $H_\t{ext}=9.14\,$kOe (b, d), and (c,e) on the external field at a fixed dc current $I =-7\,$mA; (b,c) - experiment, (d,e) - micromagnetic simulations showing the oscillations of two modes simultaneously. Dashed lines show the location of the oscillation modes in the experiment and the intensity scale indicates the normalized power (in fW/GHz for the experimental data).}
 \label{fig:1}
\end{figure}

The studied STNO is a circular nanopillar with diameter of 250 nm consisting of (Cu60$|$Py15$|$Cu10$|$Py4$|$Au25) layers (numbers indicate the thickness in nanometers), fabricated by electron beam lithography, in which top and bottom Au and Cu electrodes were designed for microwave transport measurements. The STNO is excited by a negative dc current $I_\t{dc}$, corresponding to an electron flow from the thin to the thick magnetic layer, which allows us to observe STT-induced microwave generation. In the measurements presented below, a bias magnetic field in the range  $H_\t{ext} = (8.5 -9.6)\,$kOe was applied at $\theta_H = 2^{\circ}$ from the sample normal (see Fig.~\ref{fig:1}(a)). This tilt is introduced in order to create a small misalignment between the static magnetizations of the layers, which is required for the appearance of resistance (and voltage) oscillations under almost circular magnetization precession; in the case of $\theta_H=0^{\circ}$ voltage oscillations vanish \cite{hamadeh2012autonomous}. However, this symmetry breaking also has significant consequences on the entire dynamics, as discussed in the following.

First, the bias current dependence of the voltage oscillation spectra under a constant magnetic field $H_\t{ext} =9.14\,$kOe is reported in Fig.~\ref{fig:1}(b). Within the current range from $-6\,$mA to $-8\,$mA, we observe two simultaneously auto-oscillating modes \cite{Simultat}, referred below as $\omega_0$ and $\omega_1$, which are split by about 1.1~GHz. With a further current increase, the generation frequency exhibits a pronounced jump and generation becomes seemingly single-mode (the second mode is hardly detectable). Also, this jump is accompanied by a substantial power increase (see intensity scale and supplementary materials \cite{suppl}). At even higher currents, the STNO generation demonstrates further frequency jumps and complex dynamics that are not considered here. Within the bimodal regime at $I_\t{dc} = (-6,-8.2)\,$mA, the frequencies of both modes demonstrate the same weak increase with the current amplitude, which is unexpected for almost perpendicularly magnetized STO, characterized by a strong nonlinear frequency shift \cite{slavin2009nonlinear}. 

Fig.~\ref{fig:1}(c) presents the frequency evolution with the strength of a magnetic field under a constant dc current $I_\t{dc}=-7\,$mA. The bimodal regime is robust to the field strength variation and is observed in the entire presented field range. The frequencies of both oscillation modes $\omega_0$ and $\omega_1$ vary almost linearly with the applied field; in almost perpendicularly magnetized STNO such dependence is not trivial as it may appear, indicating a significant nonlinear damping effect \cite{slavin2009nonlinear}. It is worth noting that in the case of perpendicular magnetization ($\theta_H = 0^\circ$), the generation characteristics of the STNO become fairly standard and none of the features described here indicating bimodal generation were observed \cite{hamadeh2012autonomous, suppl}.  

To get a deeper insight into the magnetization dynamics, micromagnetic simulations were performed using the FastMag software package \cite{chang2011publisher} that is based on the finite elements method (FEM). FastMag allows computing the full system including interactions between the reference and free layer via mutual spin transfer torque effects. FastMag also calculates the magneto-resistance response that can be directly compared to the experimental measurements. Material parameters were taken from the experimental study of passive dynamics (i.e., below auto-oscillation threshold) of the nanopillar \cite{naletov2011identification} (see details in supplementary \cite{suppl}). The FEM mesh edge length was chosen as 4 nm. The data evaluation of simulation was done through Aithericon \cite{Aithericon}. 

To excite the oscillation dynamics, a negative dc current was applied through the structure. The results of these micromagnetic simulations, shown in Fig.~\ref{fig:1}(d-e), reproduce the experimental measurements with a good qualitative and quantitative agreement, a clear indication that the advanced micromagnetic modeling is able to capture the features leading to the observed complex behavior. Reference simulations performed for the case $\theta_H = 0^\circ$ show standard single-mode STNO behavior, in accordance with previous experimental study \cite{hamadeh2012autonomous}.

\begin{figure}
\centerline{\includegraphics[width=\columnwidth ]{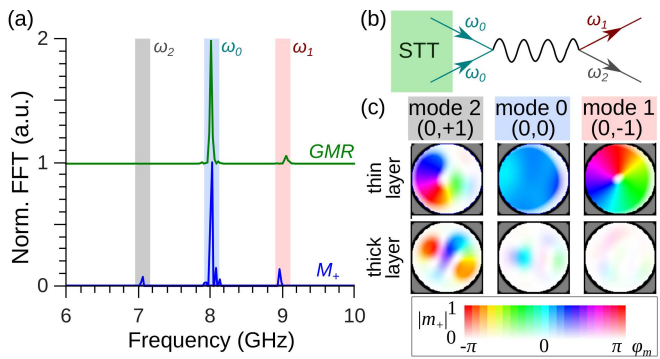}}
\caption{(a) Frequency spectrum of the $M_+$ magnetization component (blue line) and spectrum of the magnetoresistance oscillations (green line, vertically offset) of the STNO under a bias current $I_\t{dc} = -7\,$mA,  obtained by micromagnetic simulations. (b) A diagram illustrating the four-magnon scattering process leading to the multimode generation in the STNO. (c) Spatially resolved magnetization precession patterns of the three modes, obtained via FFT of magnetization time traces. The brightness indicates the amplitude and the hue indicates the phase.}
\label{fig:2}
\end{figure}

Next, we investigate spectra of the  magnetization dynamics by evaluating the fast Fourier transform (FFT) of the circularly polarized transverse magnetization components $M_+=M_x + i M_y$ emitted by the electrically driven STNO, where $i$ is the imaginary unit. In contrast to the resistance oscillation spectra, they demonstrate \emph{three} peaks -- an additional peak, referred to as $\omega_2$, is visible below $\omega_0$ (Fig.~\ref{fig:2}). The frequency of the peaks satisfies the relation $2\omega_0 = \omega_1 + \omega_2$, which leads us to the assumption that four-magnon interaction between these modes, schematically shown in Fig.~\ref{fig:2}(b), is involved in STNO dynamics. Using a mesh resolved FFT, we obtain the spatial profiles of the excitations corresponding to these peaks.The main mode $\omega_0$ is a quasi-uniform mode characterized by respectively a radial index $m=0$ and an azimuthal index $\ell=0$, which we label $(m=0,\ell=0)$ mode. This mode can be identified as the lowest-frequency mode of the isolated free layer. The peak at $\omega_1$ is the $(0,-1)$ azimuthal mode, also mostly localized within the free layer. The excitation at $\omega_2$ resembles the $(0,+1)$ azimuthal mode. However, in contrast to other excitations, this mode has comparable oscillation amplitude in \emph{both} thin and thick STNO layers indicating a strong hybridization.

\begin{figure}
 \includegraphics[width=\columnwidth]{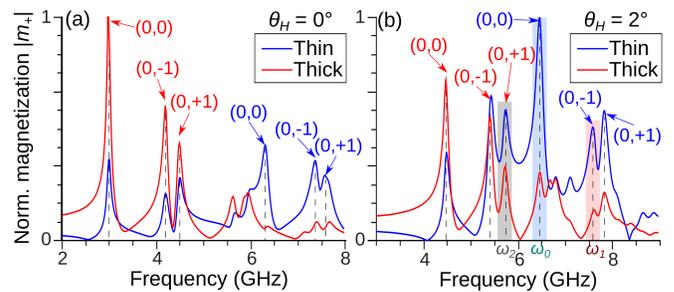}\\
\caption{Spectra of linear (low-amplitude) magnetization oscillations of the STNO nanopillar, excited by spatially nonuniform magnetic field (micromagnetic simulations); the modulus of averaged in a quarter of thin and thick layers dynamic magnetization $\vec m_+$, normalized to maximal value, is shown.}\label{fig:3}
\end{figure}

To elucidate the nature of the signal at $\omega_2$, we performed micromagnetic simulations of the STNO modes in the passive regime (no dc current) \cite{suppl}. We found that in our device, the $\ell = \pm1$ azimuthal eigenmodes of the thick layer are located below the fundamental mode of the thin layer (because of different saturation magnetization, see supplement) and these modes experience strong hybridization so that the oscillation amplitude is comparable in both layers (Fig.~\ref{fig:3}(b)). The last fact explains why the  mode at $\omega_2$ is not visible in GMR spectra: the relative angle between the layers magnetizations is only weakly changed when it is excited. Thus, we attribute the multimode generation to a four-magnon process involving the fundamental mode (0,0) of the thin layer, the azimuthal $(0,-1)$ mode of the thin layer and the  $(0,+1)$ mode of the thick layer that is strongly hybridized to the thin layer. This process satisfies the momentum conservation rule $2\ell_0 = \ell_1 + \ell_2$ and the mode eigenfrequencies are reasonably close to the energy conservation rule $2\omega_0 = \omega_1 + \omega_2$. An additional check that the thick layer mode is involved in the multimode generation was made by simulating the active STNO with a drastically increased local field in the thick layer. In this case, the thick layer modes are shifted to much higher frequencies and the above-mentioned four-magnon process becomes impossible. Indeed, no multimode generation was observed in this case. In the reference case of a perpendicular field, the thick layer modes shift to lower frequencies and thus experience much less hybridization (Fig.~\ref{fig:3}(a)), which is crucial for the multimode generation as discussed below.

In the following, we consider an analytical model to describe the dynamics of STNO modes including parametric coupling. As before, we denote the fundamental mode as mode ``0'' having the amplitude $a_0$, the $(0,-1)$ mode of the thin layer and $(0,+1)$ thick layer mode as modes 1 and 2, respectively, and consider dynamics of only these three modes. The mentioned four-magnon process is described by the term $\mathcal H^{(4)} = W_{00,12}a_1 a_2 a_0^* a_0^* + \t{c.c.}$ in the Hamiltonian of the system. Taking it into account, the dynamics of the coupled modes is described by equations
\begin{equation}\label{e:sys}
 \begin{split}
   \dot{a}_0 &+ i \tilde\omega_0 a_0 + \tilde\Gamma_0 a_0 = -2 i W_{00,12} a_0^* a_1 a_2\,, \\
   \dot{a}_1 &+ i \tilde\omega_1 a_1 + \tilde\Gamma_1 a_1 =  -i W_{00,12}^* a_0^2 a_2^* \,, \\
   \dot{a}_2 &+ i \tilde\omega_2 a_2 + \tilde\Gamma_2 a_2 =  -i W_{00,12}^* a_0^2 a_1^* \ .
 \end{split}
\end{equation}
Here, an overdot denotes time derivative, $\dot{a} = \partial_t a$, $\tilde\omega_i$ and $\tilde \Gamma_i$ are frequency and total damping of $i$-th mode accounting for nonlinear contributions. In a general case, ``total'' frequency includes linear eigenfrequency, self and cross-nonlinear shifts from all modes: $\tilde\omega_i = \omega_i + T_i |a_i|^2 + 2 \sum_{j\neq i}T_{ij} |a_j|^2$. The total damping accounts for nonlinear changes of both the damping and the STT anti-damping torque \cite{slavin2009nonlinear}: $\tilde\Gamma_i = \Gamma_i \left(1+\xi_i |a_i|^2 + 2 \sum_{j\neq i} \xi_{ij} |a_j|^2 \right) - \sigma_i I \left(1 -\tilde\xi_i |a_i|^2  - 2\sum_{j\neq i} \tilde\xi_{ij} |a_j|^2 \right)$, where $\Gamma_i = \alpha_G \omega_i$ is the linear Gilbert damping and $\sigma$ is the STT efficiency and  $\tilde\xi_{ij}$ is the nonlinear coefficients (see supplement).

Since the energy of mode 2 is more concentrated in the thick layer, while modes 0 and 1 are concentrated in the thin one, we neglect all related frequency and damping cross shifts, such as $T_{i2} = 0$, $\xi_{i2} = 0$, etc. Only for STT term ($\tilde\xi_{2i}$), such simplification is not appropriate since it is inversely proportional to the layer thickness, so the thin layer contribution to the total STT could be dominant even for mode 2. Linear frequencies for the model were extracted from micromagnetic data and nonlinear coefficients were estimated using the vector Hamiltonian formalism \cite{Tyber_ArXiv_VHF, Verba_PRB2021}, as described in Supplementary materials \cite{suppl}.

When the current increases, mode 0 first turns to the self-oscillation regime; this happens at $I_\t{th} = \Gamma_0/\sigma$ (corresponding to -6~mA in the experiment). Formal thresholds of ``isolated'' modes 1 and 2 $\Gamma_i/\sigma_i$ are larger, but competition for common STT pumping prohibits excitation of higher-order modes above their formal thresholds in the absence of other coupling \cite{Slobodianiuk_CMP2014}, and the only source for their excitation is the parametric instability of the mode~0. Then, assuming negligible amplitudes of the modes 1 and 2, we can calculate the ``virtual'' amplitude of mode 0 above the threshold in the absence of other modes, $|a_0|^2 = (\sigma I - \Gamma_0)/(\xi_0 (\sigma I - \Gamma_0))$ \cite{slavin2009nonlinear}, and the threshold amplitude $a_\t{th}$ of mode 0, above which parametric instability develops, \cite{Verba_PRB2021}
\begin{equation}
 |a_{\t{th}}|^2 = \frac{\sqrt{\tilde\Gamma_1\tilde\Gamma_2}}{|W_{00,12}|} \sqrt{1+ \frac{(2\tilde\omega_0 - \tilde\omega_1 - \tilde\omega_2)^2}{(\tilde\Gamma_1+\tilde\Gamma_2)^2}} \ .
\end{equation}
Here, the nonlinear frequency and the damping are calculated for the free-running amplitude of mode 0 at a given current and vanishing amplitudes of other modes. If $|a_0| > |a_\t{th}|$, then the parametric instability occurs and modes 1 and 2 are excited.

\begin{figure}
 \includegraphics[width=\columnwidth]{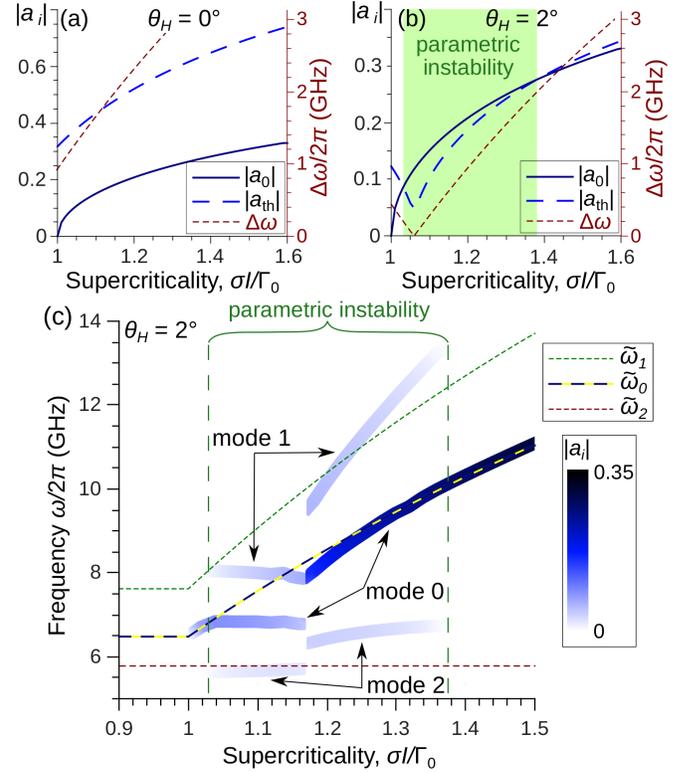}\\
\caption{(a,b) Free-running amplitude $a_0$ of the mode 0  in the absence of parametric coupling, threshold of the parametric instability $a_\t{th}$ (both - left axis), and detuning from parametric resonance condition $\Delta\omega$ (right axis) in the case of perpendicular (a) and tilted (b) field. (c) Current dependence of stationary frequency and amplitude (color-coded) of the interacting modes, obtained as numerical integration of model Eq.~(\ref{e:sys}) for $\theta_H = 2^\circ$; dashed lines show free-running frequency $\tilde\omega_0$ and eigenfrequencies $\tilde\omega_1$, $\tilde\omega_2$ which would be observed without parametric coupling.}\label{fig:4}
\end{figure} 

Dependencies of the isolated free-running amplitude $a_0$ and the parametric instability threshold are shown in Fig.~\ref{fig:4}(a,b). In the case of $\theta_H = 2^\circ$, the threshold is overcome in a certain range of currents, and a three-mode generation is expected. In contrast, in the reference case of a perpendicular field, the instability threshold is never overcome, and single-mode generation is expected. There are two reasons for this difference. The first one is the larger detuning from exact parametric resonance $\Delta\omega = 2\tilde\omega_0 - \tilde\omega_1 - \tilde\omega_2$ (see Fig.~\ref{fig:4}(a,b)) because of the different positions of the thick layer eigenmodes. However, this difference is not drastic and the second reason, namely the significantly (threefold) reduced parametric coupling $W_{00,12}$ also plays a crucial role. The latter is a result of the much lower hybridization of the modes, which is clear from the comparison of the oscillation amplitude in the thin layer at the position of $(0,+1)$ mode of the thick layer in Fig.~\ref{fig:3} (see calculation details in supplementary materials \cite{suppl}). The same reason determines the explanation for the observation of this particular pair of parametric modes at $\theta_H = 2^\circ$. Another pair, which also satisfies the momentum conservation ($(0,-1)$ thick layer mode and $(0,+1)$ thin layer mode), exhibits smaller parametric coupling because of lower hybridization of the $(0,-1)$ thick layer mode (Fig.~\ref{fig:3}(b)).

Finally, we numerically integrate Eqs.~\ref{e:sys} and present the stationary mode frequencies and amplitudes in Fig.~\ref{fig:4}(c). Starting from the threshold $\sigma I = \Gamma_0$, the amplitude and frequency of the mode 0 increase and reach the parametric instability threshold ($\sigma I/ \Gamma_0 \approx 1.035$). Above this instability threshold, all three modes are excited and their frequencies are almost frozen (a feature that could be interesting for certain applications). In this regime, the parametric process dominates and the modes have stationary amplitudes that almost satisfy the parametric resonance condition $2\tilde\omega_0 (a_i) \approx \tilde\omega_1 (a_i) + \tilde\omega_2(a_i)$. Above $\sigma I/\Gamma_0 \approx 1.18$ this generation regime cannot be sustained anymore, and we observed an abrupt jump of the mode frequencies and amplitudes as discerned in the experiment (see Fig.~\ref{fig:1}(b)). Mode 0 acquires an amplitude and a frequency that are close to the ones in the absence of parametric pumping, while modes 1 and 2 oscillate far from their ``nonlinear eigenfrequencies'' (i.e., the parametric process becomes strongly nonresonant). Their amplitudes decrease up to the current $\sigma I/\Gamma_0 \approx 1.36$, above which the instability threshold becomes too high and the STNO returns to single-mode regime. Within the multimode generation range, the amplitude of parametric mode 1 is evidently larger than of the mode 2. This can be attributed to the substantially larger total damping rate of the mode 2. This circumstance also complicates the experimental observation of  mode 2.

Overall, the model results are very similar to experimental observations. The finite range of multimode generation, the small (vanishing) frequency slope, the almost constant oscillation amplitude in the multimode regime, and the frequency jump are all observed in experiment.  There is a quantitative difference in mode oscillation frequencies, which is on account of (i) the approximate calculation of nonlinear coefficients, particularly the phenomenological coefficient of nonlinear damping, and (ii) neglecting of the Oersted fields in the model that, according to experimental measurements below the threshold, result in the mode frequency increase about $70\,\t{MHz}/\t{mA}$, almost the same for different modes.  

To conclude, we found by experiment a regime of simultaneous multimode microwave emission by a nanopillar-based STO. The physical process behind this is a second-order parametric instability between a pair of radially symmetric magnon modes and a pair of opposite azimuthal index modes, without any spatio-temporal overlap. The process, however, satisfies the constraint of both angular momentum and energy conservation through the hybridization of eigenmodes between the thin and the thick layer. This co-generation has the additional feature of being incommensurable. The splitting between modes can thus be tuned by changing the characteristics of the nano-pillar such as its diameter. A future work direction will be to exploit non-uniform magnetic textures as a means to control continuously the splitting by an external bias parameter, such as the magnetic field. A particularly promising candidate is the vortex, where the splitting between the $\ell=\pm 1$ has been shown to vary strongly as a function of $H_0$. We believe that the simultaneous self-generation of multiple frequencies in dc-driven spintronic elements has large potential for applications using frequency multiplexing techniques in neuromorphic and wave-based approaches.

\begin{acknowledgements}

This work has been  supported  by the European Research Council within the Starting Grant No. 101042439 "CoSpiN" and by the Deutsche Forschungsgemeinschaft (DFG, German Research Foundation) - TRR 173 - 268565370" (project B01) and by the French Grants ANR-21-CE24-0031 Harmony and the EU-project H2020-2020-FETOPEN k-NET-899646 and  by the National Academy of Sciences of Ukraine, project \#0122U002462 and  the project PRIN 2020LWPKH7 funded by the Italian Ministry of University and Research.
\end{acknowledgements}

\bibliography{ref}

\begin{thebibliography}{40}
\bibitem [{Sim()}]{Simultat}%
  \BibitemOpen
  \href@noop {} {}\bibinfo {note} {We have checked it experimentally by
  performing a synchronization experiment. It showed that both $\omega_0$ and
  $\omega_1$ are locked together when an external oscillation is additionally
  applied at either $\omega_0$ or $\omega_1$. We interpret this as a proof of
  the simultaneity of the emission.}\BibitemShut {Stop}%
\end{thebibliography}


\end{document}